\documentclass[12pt]{article}
\usepackage{graphicx}
\usepackage{epsfig,graphicx}
\usepackage{amsmath,amssymb}

\title
{\bf Seesaw model in SO(10) with an upper limit 
on right-handed neutrino masses}

\author
{\bf M. Abud, F. Buccella,  D. Falcone\\Dipartimento di Scienze Fisiche, Universit\`a di Napoli\\and
INFN, Sezione di Napoli, Italy\\
\bf L. Oliver \\ Laboratoire de Physique Th\'eorique, Universit\'e de Paris XI, Orsay Cedex, France}
\date{}

\newcommand{\beq}{\begin{eqnarray}}
\newcommand{\eeq}{\end{eqnarray}}
\newcommand{\be}{\begin{equation}}
\newcommand{\ee}{\end{equation}}

\newcommand{{\SD}}{\rm SD}

\newcommand{{\Mc}}{\mathcal{M}}

\begin{document}

\maketitle

\vskip 0.5 truecm

\begin{abstract} 

\vskip 0.5 truecm

In the framework of $SO(10)$ gauge unification and the seesaw 
mechanism, we show that the upper bound on the mass of the heaviest 
right-handed neutrino $M_{R_3} < 3 \times 10^{11}$ GeV, given by the 
Pati-Salam intermediate scale of $B-L$ spontaneous symmetry breaking,
constrains the observables related to the left-handed light neutrino mass matrix. 
We assume such an upper limit on the masses of right-handed neutrinos
and, as a first approximation, a Cabibbo form for the matrix $V^L$ that diagonalizes the 
Dirac neutrino matrix $m_D$. 
Using the inverse seesaw formula, we show that our hypotheses imply a triangular relation 
in the complex plane of the light neutrino masses with the Majorana phases. 
We obtain normal hierarchy with an absolute scale for the light neutrino spectrum.
Two regions are allowed for the lightest neutrino mass $m_1$ and for the Majorana phases,
implying predictions for the neutrino mass measured in Tritium decay and for the double beta 
decay effective mass $|<m_{ee}>|$. 
 
\end{abstract}

\vspace{1cm} DSF/8/2011\par
\vspace{0.5cm} 
LPT-Orsay 11-96

\newpage \pagestyle{plain}

\section{Introduction}

The present status of neutrino oscillations, conceived many years ago 
by Pontecorvo \cite{P}, provides the following approximate values for the square mass 
differences and the  mixing angles of the PMNS matrix \cite{MNS,BP} :
\beq
\Delta m_s^2 = |m_2|^2 - |m_1|^2 \simeq 8 \times 10^{-5}\ \rm{eV}^2 \\
\tan^2 \theta_s \simeq 0.4\\
\Delta m_a^2 = |m_3|^2 - \textrm{cos}^2\theta_s\ |m_2|^2 
- \textrm{sin}^2 \theta_s\ |m_1|^2  \simeq 2.5 \times 10^{-3}\ \rm{eV}^2 \\
\tan^2 \theta_a \simeq 1
\label{1}
\eeq
The following experimental limits constrain the effective mass matrix
of the left-handed neutrinos :
\beq
 2.6 \times 10^{-3}\ \textrm{eV} < m_{\nu_e} <  2.2\ \textrm{eV} \\ 
|<m_{ee}>|  < 0.4\ \rm{eV} \\ 
0.06\ \textrm{eV} < \sum_i m_i < 0.6\ \textrm{eV}.\label{eq:Bounds}
\label{2}
\eeq

These upper limits are respectively obtained from the high energy spectrum 
of the electron in nuclear beta decay, from the upper limit on the rate in 
neutrinoless double beta decay (for Majorana neutrinos) and from cosmology.\par
The lower limits on $m_{\nu_e}$ and $\sum_i m_{i}$ are respectively obtained 
from the bounds
\beq
m_{\nu_e} > \Delta m_s \sin^2\theta_s \\ 
\sum_i m_{i} > \Delta m_s + \Delta m_a
\label{2bis}
\eeq

An upper limit has also been found for the component of the $\nu_{eL}$ along 
the third mass eigenstate, supposedly the heaviest, i.e. the one that is not 
involved in solar neutrino oscillations :
\beq
\sin^2 \theta_{13} < 0.03
\label{3}
\eeq

It is generally recognized that $SO(10)$ unified gauge theories 
\cite{SO10} provide a very natural framework for the seesaw model \cite{SS}, 
accounting naturally  for the fact that left-handed neutrinos have masses 
several orders of magnitude smaller than the charged fundamental fermions.
Indeed, the {\bf 16} representation of $SO(10)$ contains, besides the  {\bf 10} and
${\bf \bar{5}}$ of $SU(5)$, a singlet that can get a large mass, unrelated to 
the electroweak symmetry breaking scale.\par
Moreover, in the most appealing gauge unified $SO(10)$ model, the one with 
$SU(4) \times SU(2) \times SU(2)$ Pati-Salam \cite{PS} intermediate 
symmetry, $B - L$ is broken around $3 \times 10^{11}$ GeV \cite{BCST,M,ACA}, 
providing the scale for right-handed neutrino masses by the $\Delta L=2$ 
vacuum expectation value (VEV) of the {\bf 126} representation.\par

In $SO(10)$ one expects a spectrum for the eigenvalues of the Dirac
neutrino mass matrix that is similar to the masses 
of the quarks with charge ${2 \over 3}$, apart from some scale factor due to 
the different scale dependence of quark and leptons masses.\par
It is also very reasonable to assume that the matrix $V^L$ appearing in
the biunitary trasformation that diagonalizes the  Dirac neutrino mass 
matrix $m_D$ has the same structure as the Cabibbo-Kobayashi-Maskawa 
matrix $V_{CKM}$ \cite{CKM}, namely
a hierarchical structure, the mixing 
angle between the first two generations being larger than the other angles.
This statement is stricly correct within the simplifying hypothesis of assuming 
that the Higgs bosons providing the Dirac masses and mixing belong to 
${\bf 10}$ representations.

\section{The inverse seesaw}

In this paper we intend to deduce the consequences of two main hypothesis :\par
(i) We assume an upper limit for the right-handed neutrino masses.\par
(ii) Within the $SO(10)$ gauge unification scheme, the Dirac mass matrix (eigenvalues and mixing)
has the same structure as the up quark mass matrix (eigenvalues and mixing).\par

More quantitatively, we shall assume for the eigenvalues of the Dirac neutrino 
mass matrix the same values than in \cite{AFS}, namely : 
\beq
\label{3-1e}
m_{D_1} = 10^{-3}\ \textrm{GeV} \qquad \qquad m_{D_2} = 0.4\ \textrm{GeV} \qquad \qquad m_{D_3} = 100\ \textrm{GeV}
\eeq

\noindent Moreover we shall take for $V^{L}$ a matrix that, to begin with, has the Cabibbo 
form with only $\theta_{12}$ different from zero
\beq
V^L = \left(
        \begin{array}{ccc}
       \cos{\theta_{12}}  & \sin{\theta_{12}} & 0 \\
        - \sin{\theta_{12}}  & \cos{\theta_{12}}  & 0 \\
         0 & 0 & 1  \\
        \end{array}
        \right)
\label{4}
\eeq
which was a very instructive approximation \cite{BFO}. The rest of the angles are
considered as perturbations relatively to the simple ansatz (\ref{4}) and, as shown
in \cite{BFO}, even the quantitative features of the light left-handed neutrino
spectrum are correctly described. 
  
Let us consider the inverse seesaw formula :
\beq
M_R = -m_D^t\ m_L^{-1}\ m_D
\label{5}
\eeq
Diagonalizing the neutrino Dirac mass matrix $m_D$ by
\beq
m_D = V^{L\dagger }\ m^{diag}_D\ V^{R}
\label{6}
\eeq
one gets the relation
\beq
M_R = -\ V^{Rt}\ m_D^{diag}\ V^{L*}\ m_L^{-1}\ V^{L\dagger }\ m_D^{diag}\ 
V^{R}
= -\ V^{Rt}\ m_D^{diag}\ A^L\ m_D^{diag}\ V^{R}. 
\label{7}
\eeq
where the matrix $A^L$ is defined by \cite{BF} :
\beq
A^L =  V^{L*}\ m_L^{-1}\ V^{L\dagger }
\label{7-1}
\eeq

Moreover, within $SO(10)$, with the electroweak Higgs boson belonging to the {\bf 10} and {\bf 126} representations, and {\it no component along the {\bf 120} representation}, the mass matrices are symmetric. As a consequence, the unitary matrices $V^R$ and $V^L$ that diagonalize Dirac neutrino matrix (\ref{6}) are related :
\beq
\label{7-2e}
V^R = V^{L*}
\eeq
and the matrix $M_R$ (\ref{7}) becomes 
\beq
M_R = -\ V^{L+}\ m_D^{diag}\ A^L\ m_D^{diag}\ V^{L*}
\label{7-3e}
\eeq

\noindent The Cabibbo limit (\ref{4}) taken by us would be a good approximation of $V^L$ in the limit of quark-lepton symmetry, with only components along the ${\bf 10}$ representations for the electroweak Higgs, where $V^L$ should be equal to $V_{CKM}$.

The neutrino mass matrix $m_L$ is diagonalized by the PMNS unitary 
neutrino mixing matrix, which reads : 
\beq
U \simeq \left(
        \begin{array}{ccc}
       c_s & s_s & 0 \\
        -\ {s_s \over \sqrt{2}} & {c_s \over \sqrt{2}} & {1 \over \sqrt{2}} \\
        {s_s \over \sqrt{2}} & -\ {c_s \over \sqrt{2}} & {1 \over \sqrt{2}} \\
        \end{array}
        \right).\ \textrm{diag}(1,e^{i\alpha},e^{i\beta})
\label{8}
\eeq

\noindent in the approximation that we will consider here for the angle (\ref{3})
\beq
\sin{\theta_{13}} \simeq 0
\label{8-1}
\eeq

\noindent In writing (\ref{8}) we have taken the maximal mixing angle for atmospheric neutrino 
oscillation and $s_s \equiv \sin{\theta_s}$ ($c_s \equiv \cos{\theta_s}$) and the angles $\alpha$ and $\beta$ are the Majorana phases.  We use in (\ref{8}) the notation of Davidson et al. \cite{DNN} for the Majorana phases, that have the ranges $0 \leq \alpha \leq \pi, 0 \leq \beta \leq\pi$. In the PDG Tables \cite{PDG} they are defined as $\alpha_{21}/2$ and $\alpha_{31}/2$, with $0 \leq \alpha_{21} \leq 2\pi, 0 \leq \alpha_{31} \leq 2\pi$.\par
Then, the left-handed neutrino light mass matrix reads
\beq
m_L = U^*\ m_L^{diag}\ U^\dagger  	\qquad \qquad 		
m_L^{-1} = U\ (m_L^{diag})^{-1}\ U^t 
\label{9}
\eeq

\noindent where
\beq
\label{9-1}
m_L^{diag} = \textrm{diag}(m_1,m_2,m_3) \qquad \qquad \qquad \qquad m_i \geq 0 \qquad (i = 1,2,3)
\eeq

\noindent are the light neutrino masses, real positive parameters, since the Majorana phases have been factorized out, as it should.

For the inverse $m_L^{-1}$ of the matrix (\ref{9}) we will have : 
\beq
m_L^{-1} =  \left(
        \begin{array}{ccc}
        {c_s^2 \over m_1} + {s_s^2 \over e^{-2i\alpha} m_2} & 
- {c_s s_s \over \sqrt{2}} \left ({1 \over m_1} 
- {1 \over e^{-2i\alpha} m_2} \right ) & {c_s s_s \over \sqrt{2}} 
\left ({1 \over m_1} - {1 \over e^{-2i\alpha} m_2} \right ) \\  
- {c_s s_s \over \sqrt{2}} \left ({1 \over m_1} 
- {1 \over e^{-2i\alpha} m_2} \right )  & {1 \over 2}  
\left ({s_s^2 \over m_1} + {c_s^2 \over e^{-2i\alpha} m_2} 
+ {1 \over e^{-2i\beta}m_3} \right ) &  - {1 \over 2}  
\left ({s_s^2 \over m_1} + {c_s^2 \over e^{-2i\alpha} m_2} - {1 \over e^{-2i\beta}m_3} \right ) \\   
{c_s s_s \over \sqrt{2}} \left ({1 \over m_1} 
- {1 \over e^{-2i\alpha} m_2} \right ) &  - {1 \over 2}  
\left ({s_s^2 \over m_1} + {c_s^2 \over e^{-2i\alpha} m_2} 
- {1 \over e^{-2i\beta}m_3} \right ) &  
{1 \over 2}  \left ({s_s^2 \over m_1} + {c_s^2 \over e^{-2i\alpha} m_2} 
+ {1 \over e^{-2i\beta}m_3} \right ) \\
        \end{array}
        \right)
\label{10}
\eeq
 \newline
Therefore, being $m_L^{-1}$ symmetric and $V^L$ unitary, the matrix $A_L$ is also symmetric.\par
Of interest for our discussion will be the consideration of the matrix $m_D^{diag} A^L m_D^{diag}$ that
enters in r.h.s. of the expression (\ref{7}) : 
\beq
 m_D^{diag} A^L m_D^{diag} = \left(        \begin{array}{ccc}
       A^L_{11} m^2_{D1} & A^L_{12} m_{D1} m_{D2} &A^L_{13} m_{D1} m_{D3} \\
       A^L_{12} m_{D1} m_{D2} & A^L_{22} m^2_{D2} & A^L_{23} m_{D2} m_{D3} \\
       A^L_{13} m_{D2} m_{D3} & A^L_{23} m_{D2} m_{D3} & A^L_{33} m^2_{D3} \\
        \end{array}
        \right)
\label{20-1}
\eeq

The coefficient $A^L_{33}$ of the square of the highest Dirac eigenvalue (\ref{3-1e}),
$m_{D_3}^2 = (100\ \rm{GeV})^2$, within the simplifying hypotheses of a Cabibbo form for $V^L$ (\ref{8})
and $s_{13} = 0$ (\ref{8-1}), is \cite{AB}\cite{BFO} :
\beq
A^L_{33} = (m^{-1}_{L})_{33}
= {1 \over 2} \left({s^2_s \over m_1} + {c^2_s \over e^{-2i\alpha} m_2} + {1 \over e^{-2i\beta}m_3}\right)
\label{11}
\eeq

\noindent and in the limit $m_{D_1}, m_{D_2} << m_{D_3}$ (\ref{3-1e}) one has roughly
\beq
M_{R_3} \sim \mid A^L_{33} \mid m^2_{D_3} = {1 \over 2} \left| {s^2_s \over m_1} + {c^2_s \over e^{-2i\alpha} m_2} + {1 \over e^{-2i\beta}m_3} \right| m^2_{D_3}
\label{11-1}
\eeq

The expression (\ref{11}) found for $A^L_{33}$ follows from the assumption (\ref{4}) for the matrix $V^L$. Let us notice that in all generality it will also depend on the square of the mixing angle between the third generation and the other two lighter ones, that is assumed to be small.

Let us first remark that a rather conservative upper limit on the mass of the heaviest right-handed neutrino of the order 
\beq
M_{R_3} \leq 10^{15}\ \rm{GeV}
\label{11-2}
\eeq

\noindent implies a lower limit for the mass of 
the lightest left-handed neutrino, since in the small $m_1$ region, when the 
first term in (\ref{11}) dominates, one should have, with the value (\ref{3-1e}) for $m_{D_3}$ : 
\beq
 {1 \over 2} {s_s^2 \over m_1} \times 10^{4}\ \rm{GeV}^{2}\le  10^{15}\ \rm{GeV}
\label{11-3}
\eeq
which implies 
\beq
 m_1\ge  1.4\times 10^{-3}\ \rm{eV} 
\label{12}
\eeq
Since $m_2$ and $m_3$, according to (1) and (3) are monotonically 
increasing  functions of $m_1$, one has 
\beq
|\textrm{det}(m_L)|\ge 1.4\times 
10^{-3} \Delta m_s \Delta m_a=  6.43 \times 10^{-7}\ \textrm{eV}^3 
\label{12-1}
\eeq

\noindent and for the Majorana mass matrix of right-handed neutrinos one has :
\beq
|\textrm{det}(M_R)| \le 2.5\times 10^{30}\ \textrm{GeV}^3
\label{12-2}
\eeq

\section{Imposing an upper bound on the heaviest $M_{\nu_{R}}$ eigenvalue}

Let us stress that large cancellations are required in (26) if we impose 
to the masses of the right-handed neutrinos the more stringent limit 
\beq
M_{R_3} \leq 3 \times 10^{11}\ \rm{GeV}
\label{12-3}
\eeq

\noindent i.e. the scale of $B - L$ spontaneous symmetry breaking
in the $SO(10)$ unified gauge theory with Pati-Salam intermediate symmetry.\par 

The trivial bound 
\beq
|A_{33}^{L}| \leq {1 \over 2} \left({s_{s}^{2} \over m_{1}}+{c_{s}^{2} \over m_{2}} + {1 \over m_{3}}\right)
\label{12-4}
\eeq

\noindent would be effective to constrain $M_{R}$ to be smaller than $10^{11}$ GeV only
in a region of {\it unrealistically large} neutrinos masses. In fact $|A_{33}^{L}|$
should be smaller about two orders of magnitude than ${1\over 2 m_{3}}$, taking into account the 
upper limit on $\Sigma_i m_i$ (7).\par

From (\ref{3-1e}) and (\ref{11-1}) we see that (\ref{12-3}) implies
\beq
|A_{33}^{L}| < 3 \times 10^{-2}\ \textrm{eV}^{-1} << 2.5\ \textrm{eV}^{-1} \leq {1 \over {2m_3}}
\label{12-6}
\eeq

More precisely, only the third term in the r.h.s. of (\ref{12-4}) would give rise at least, by assuming  the largest value
for $m_3$ consistent with the square masses differences fixed by the 
oscillations (3) and the rather conservative cosmological limit (7)
on the sum of their masses, $0.6$ eV, to a mass around $M_{R_3} \simeq 2.5 \times 10^{13}$ GeV,
two orders of magnitude larger than the value expected in the ordinary $SO(10)$
unified model with Pati-Salam intermediate symmetry.\par
Therefore, we underline again that one needs a strong cancellation between the three terms in (\ref{11}), which have moduli
related by the square mass differences implied by neutrino oscillations.

Notice the very important point that this is already a hint for large relative Majorana phases. In this respect, it is interesting to look for the implications for the neutrinoless double 
beta decay effective mass (6) :
\beq
<m_{ee}>\ = c_s^2\ m_1 + s_s^2\ e^{-2i\alpha} m_2
\label{13}
\eeq

\noindent Owing to (\ref{11}) $<m_{ee}>$ can be exactly expressed in terms of $m_{i}\ (i=1,2,3)$ and $A_{33}^{L}$ by the formula
\beq
<m_{ee}>\ = - e^{-2i(\alpha-\beta)}\ \frac{m_1 m_2}{m_3} + 2 A^L_{33} e^{-2i\alpha} m_1 m_2
\label{13-1}
\eeq

\noindent Taking into account (\ref{12-6}), one can neglect the second term in the r.h.s. of (\ref{13-1}) and we obtain, just from the imposed upper limit on $M_{R_3}$ (\ref{12-3}), the simple expression for $<m_{ee}>$ :
\beq
<m_{ee}>\ = - e^{-2i(\alpha-\beta)}\ \frac{m_1 m_2}{m_3}
\label{13-2}
\eeq

In the following we shall take 
\beq
A^L_{33} = {1 \over 2} \left({s^2_s \over m_1} + {c^2_s \over e^{-2i\alpha} m_2} 
+ {1 \over e^{-2i\beta} m_3} \right) = 0
\label{14}
\eeq
since the second term in the r.h.s. of (\ref{13-1}) is at most $1\%$ of the first one.
Notice that relation (\ref{14}) follows from the fact that $A^L_{33}$, due to eqn. (\ref{20-1}), is affected by the square of the largest mass $m_{D_3}$ and has nothing to do with the values of the other eigenvalues in (\ref{3-1e}). 

\section{A triangle in the complex plane of light neutrino masses and Majorana phases}

Let us now examine carefully the consequences of the condition (\ref{14}). This cancellation condition defines {\it a triangle in the complex plane} :
\beq
{s^2_s \over m_1} + {c^2_s \over e^{-2i\alpha} m_2} + {1 \over e^{-2i\beta} m_3} = 0
\label{14-1}
\eeq

\noindent that we have drawn in Fig 1. 

\begin{tabular}{c}
\includegraphics[scale=0.9]{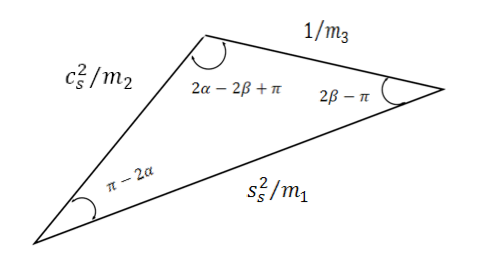} \\ Fig. 1. The triangle in the complex plane (\ref{14-1}). The sides are given in terms of the inverses of \\ \qquad \qquad the light neutrino masses and the angles as functions of the Majorana phases $\alpha$ and $\beta$ \\ \qquad of the light neutrino mixing matrix (\ref{8}). \ \ \ \qquad \qquad \qquad \qquad \qquad \qquad \qquad \qquad \qquad \\
\\\end{tabular}

Eqns. (1) and (3) give $m_2$ and $m_3$ in terms of $m_1$ and (\ref{14-1}) may be satisfied if one has the inequality

\beq
\left|{s^2_s \over m_1}-{c^2_s \over m_2}\right|\leq  {1 \over m_3} \leq  {s^2_s \over m_1}+{c^2_s \over m_2}
\label{14-1ter}
\eeq

\noindent which is violated for $m_1 < 2.9926 \times 10^{-3}$ eV or in the range ($6.2194 \times 10^{-3}$ eV, $1.9861 \times 10^{-2}$ eV).\par 

We thus get two regions where the triangular relation holds :
\beq
{\rm{Region}\ I} \qquad \qquad  r_{1} \leq m_{1} \leq r_{2} \qquad \qquad (r_{1} = 2.9926\times 10^{-3}\ {\rm{eV}},\ r_{2} = 6.2194\times 10^{-3}\ {\rm{eV}}) \qquad
\label{14-2}
\eeq
\beq
{\rm{Region}\ II} \qquad \qquad m_{1} \geq r_{3} \qquad \qquad \qquad (r_{3} = 1.9861\times 10^{-2}\ {\rm{eV}}) \qquad \qquad \qquad \qquad \qquad \qquad
\label{14-3}
\eeq

On the boundaries of both regions $m_{1}=r_{1}, r_{2}, r_{3}$ one has $\sin(2\alpha) = \sin(2\beta) = 0$ but these two quantities can be reasonably large in their interior.\par 
We plot in Figures 2 and 3 the dependence of the Majorana phases $\alpha$ and $\beta$ for both regions I and II as functions of $m_1$ (in $10^{-3}$ eV units).\par

\begin{tabular}{c}
\qquad \qquad \includegraphics[scale=0.65]{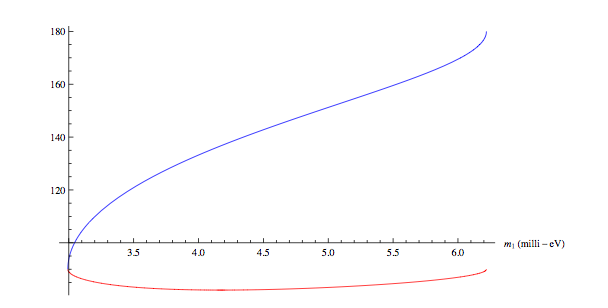} \\ Fig. 2. The Majorana phases $\alpha$ (red) and $\beta$ (blue) in Region I as function of  $m_1$ in $10^{-3}$ eV units\\
\\\end{tabular}

\begin{tabular}{c}
\qquad \qquad \includegraphics[scale=0.65]{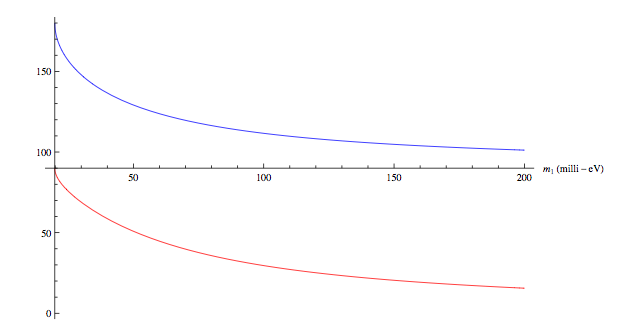} \\ Fig. 3. The Majorana phases $\alpha$ (red) and $\beta$ (blue) in Region II as function of  $m_1$ in $10^{-3}$ eV units\\
\\\end{tabular}

As we see in Fig. 2, in Region I the phase $\alpha$ decreases from $90^{o}$ to $82.1^{o}$ at $m_{1} = 4.1\times 10^{-3}\ \rm{eV}$ and gets back to $90^{o}$ at $m_1 = r_2$, while $\beta$ increases from $90^{o}$ to $180^{o}$ (notice that we take $\alpha$ in the range ($0^{o},90^{o}$) and $\beta$ in the range ($90^{o},180^{o}$), but of course (\ref{14-1}) holds also with the opposite choice). As we will see below, the fact that $\alpha$ is rather close to $90^{o}$ for Region I will imply a strong cancellation in the effective mass $<m_{ee}>$ of neutrinoless double beta decay.\par
In Region II both Majorana phases can get large values for moderate values of $m_{1}$, where the sides of the triangle (\ref{14-1}) are not very different.\par 

Notice now the important remark that for both regions one must have {\it normal hierarchy}. The reason is the following. From eqn. (\ref{14-1}) one gets the relation
\beq
\tan^2{\theta_s} = - \frac{m_1(e^{-2i\alpha} m_2 + e^{-2i\beta} m_3)}{e^{-2i\alpha} m_2(m_1 + e^{-2i\beta} m_3)}
\label{15}
\eeq 

\noindent which, for the inverted hierarchy ($m_3 << m_1, m_2$), would be about -1.\par
For the normal hierarchy ($m_1, m_2 << m_3$), eqn. (\ref{15}) becomes $\tan^2{\theta_s} = - \frac{m_1}{e^{-2i\alpha} m_2}$ and to have $\tan^2{\theta_s} \simeq 0.4$ (eqn. (2)) one needs
\beq
{m_1 \over m_2} \simeq 0.4  \qquad \qquad \qquad \qquad \alpha \simeq {\pi \over 2} 
\label{14-6}
\eeq

As we have seen above, when $A^L_{33}$ vanishes, one has, from (\ref{13-2}) 
\beq
|<m_{ee}>|\ \simeq  \frac{m_1 m_2}{m_3} 
\label{16}
\eeq
so that, once  $m_1$ is fixed, the three quantities in (\ref{14})
are also fixed, with  $m_2$ and $m_3$ given by (1) and (3).\par
In the present scheme we have therefore for $|<m_{ee}>|$ the appealing expression (\ref{16}), which implies a negative interference between the two terms in (\ref{13}) for small $m_1$, and a positive one when $m_1$ approaches the cosmological bound.   

On the other hand, the mass $m_{\nu_e}$ can be obtained from
\beq
m_{\nu_e} = 
c_s^2 m_1 + s_s^2 m_2
\label{17}
\eeq

\subsection {A further discussion on the constraint $M_{R_3} < 3 \times 10^{11}$ GeV}

Besides the main constraint (\ref{14-1}), some words of caution are necessary to prevent a mass for the heavier right-handed 
neutrino $M_{R_3}$ to be not larger than $3 \times 10^{11}$ GeV. We have also to check that 
\beq
|A_{23}^{L}| \leq 7.5\ {\rm{eV}}^{-1} 
\label{17-1}
\eeq

\noindent because $A_{23}^{L}$ multiplies the product of the two highest 
eigenvalues of the Dirac matrix, as we can see in (\ref{20-1}), 
$m_{D_2}m_{D_3} \sim 40\ \rm{GeV}^2$. It depends on the $\theta_{12}$ 
mixing angle and is given by
$$
A_{23}^{L} =-{1 \over 2} \left({s^2_s \over m_1} + {c^2_s \over m_2 e^{-2i\alpha}} - {1 \over m_3 e^{-2i\beta}} \right) \cos{\theta_{12}}
  - {1 \over \sqrt{2}}\ c_s s_s \left({1 \over m_1} - {1 \over m_2 e^{-2i\alpha}} \right) \sin{\theta_{12}}
$$
\beq  
= {1 \over m_3 e^{-2i\beta}} \cos{\theta_{12}} - {1 \over \sqrt{2}}\ c_s s_s \left({1 \over m_1} - {1 \over m_2 e^{-2i\alpha}} \right) \sin{\theta_{12}}
\label{18}
\eeq
At the boundary $m_1= r_{1}, r_{2}, r_{3}$ 
of the allowed regions, we can tune the value of $\theta_{12}$ 
in order that  $A_{23}^{L}=0$ holds, as following\ :
\beq
\tan{\theta_{12}} = \sqrt{2}\ {1 \over c_s s_s} {m_1 m_2e^{-2 i \alpha} \over m_3e^{-2 i \beta}(m_2e^{-2 i \alpha} - m_1)}
\label{19}
\eeq

\noindent implying $\tan{\theta_{12}} = 0.14, 0.24\ {\rm{and}}\ 0.6$, at 
$m_1 = r_1, r_2\ {\rm{and}}\ r_3$ respectively, where $\sin(2\alpha) = \sin(2\beta) = 0$, as we have seen above. In the first region, as soon 
as in the complex plane ${1 \over m_3 e^{-2i\beta}}$ forms a large angle with ${1 \over m_1} - {1 \over m_2 e^{-2i\alpha}}$, 
the cancellation between the two terms in (\ref{18}) is impossible and, when they are 
just orthogonal, the coefficient of the term proportional to 
$m_{D_2}m_{D_3}$ is at least of order ${1 \over m_3 e^{-2i\beta}}$, giving rise to two right-handed 
neutrinos around $8 \times 10^{11}$ GeV and a lowest state around $0.32\times 10^{6}$ GeV.\par

In order to avoid a too small value for the mass of the lightest right-handed 
neutrino, a necessary condition is that $|A^L_{22} A^L_{33} - (A^L_{23})^2|$ is smaller than 
$|(A^L_{23})^2|$. This can be obtained by relaxing the condition $A^L_{33} = 0$. 
However, one gets anyway a too small mass for the lightest right-handed neutrino because
of the range allowed for
$$A^L_{22} = \frac{\cos^2{\theta_{12}}}{2} 
\left(\frac{s^2_s}{m_1} + \frac{c^2_s}{m_2 e^{-2i\alpha}} + \frac{1}{m_3 e^{-2i\beta}}\right)$$
\beq
+\sqrt{2}\ \sin{\theta_{12}} \cos{\theta_{12}}\ c_s s_s 
\left(\frac{1}{m_1}-\frac{1}{m_2 e^{-2i\alpha}}\right) 
+\sin^2{\theta_{12}} \left(\frac{c^2_s}{m_1} + \frac{s^2_s}{m_2 e^{-2i\alpha}}\right) 
\label{21}
\eeq

\noindent $A^L_{22}$ would be equal to $A_{33}^L$ in the limit of vanishing $\theta_{12}$.
So, near the boundaries of Region I one gets the choice made recently \cite{BFO} of a compact 
neutrino spectrum, as it is also the case with a large value of $\tan\theta_{12}$ near $r_3$.
The other values of $m_1$ consistent with eqn. (\ref{13-2}) imply a value higher than $3 \times 10^{11}$ GeV 
for the two heaviest right-handed neutrinos, and a small value for the lightest one.

\section{Phenomenological implications for low-energy $\nu_L$ physics}

In conclusion, the choice of a compact spectrum seems the most natural,
but it is useful to describe the phenomenological consequences of the 
other scenarios.
We shall write the phenomenological consequences for the quantities,
for which there are the limits written in (5)-(7) for the two regions (\ref{14-2}) 
and (\ref{14-3}) in the triangle (\ref{14-1}).
For the sum of the moduli of the neutrino masses we find in Region I 
values slightly above the lower limit $|\Delta m_a|$ + $|\Delta m_s| \geq 0.06$ eV, 
while in Region II the sum of the neutrino masses is at least
$0.96 \times 10^{-1}$ eV, it grows almost linearly and saturates the bound 
at $m_1=0.198$ eV.\par
We get always a small value for $|<m_{ee}>|$, in the range $(5.6 \times 
10^{-4} -1.3\times 10^{-3})$ eV in Region I, while 
in Region II the relevant range is $(8.5 \times 10^{-3}-0.2)$ eV. We 
have limited the evaluation in Region II to $m_1\le 0.2$ eV, 
according to the bound (7).\par
For $m_{\nu_e}$, the neutrino mass intervening in the tritium 
decay, it is confined to the ranges $(4.8 - 7.5)\times 10^{-3}$ eV for 
Region I and $(2\times 10^{-2} - 0.2)$ eV for Region II.\par

To summarize, we obtain the following numerical results :\par
$${\rm{Region}\ I}$$
\beq
m_{\nu_e} = (4.8 - 7.5)\times 10^{-3}\ {\rm{eV}} \qquad \ \  \sum_i m_{i} = 0.1\ {\rm{eV}} \qquad \ \  |<m_{ee}>| = (0.6  -1.3) \times 10^{-3}\ {\rm{eV}}
\label{22-1}
\eeq
$${\rm{Region}\ II}$$
\beq
m_{\nu_e} = (2\times 10^{-2} - 0.2)\ {\rm{eV}} \qquad \sum_i m_{i} = (0.1 - 0.6)\ {\rm{eV}} \qquad |<m_{ee}>| = (8.5 \times 10^{-3}-0.2)\ {\rm{eV}}
\label{22-2}
\eeq

\section{Conclusions}

With reasonable hypotheses in the framework of $SO(10)$
unified theories, and by imposing the simple assumption of an upper bound on the mass of the heaviest 
right-handed neutrino $M_{R_3} < 3 \times 10^{11}$ GeV, as suggested by
a Pati-Salam intermediate scale of $B-L$ spontaneous symmetry breaking, 
one gets interesting predictions for the physical quantities related to the 
effective mass matrix of the light left-handed neutrinos, namely on the mass of the lightest neutrino and on the Majorana phases.\par
Using the inverse seesaw formula, we have shown that our hypothesis of an upper bound for the right handed neutrino masses implies a triangular relation in the complex plane of the light neutrino masses with the Majorana phases. In a straightforward way we thus have predicted, on the one hand, normal hierarchy for the light neutrinos and a lower limit and an exclusion region for the mass of the lightest left-handed neutrino $m_1$, implying an absolute scale for the light neutrino spectrum.\par 
The allowed regions for $m_1$ are the 
range $m_1 = (3.0 - 6.2) \times 10^{-3}$ eV and the lower bound $m_1 \geq 2.0 \times 10^{-2}$ eV.
For small $m_1$, one of the Majorana phases can be close to ${\pi \over 2}$, and we get a strong cancellation in the effective mass 
$|<m_{ee}>|$ of neutrinoless double beta decay, and for light neutrino masses near the cosmological bound we obtain a positive interference for this quantity. Within our scheme we obtain also an interesting formula for $|<m_{ee}>|$ just in terms of the three light neutrino masses, that is valid in both domains allowed for $m_1$.\par

\newpage

\noindent {\large \bf Acknowledgements}

\par \vskip 0.5 truecm

It is a pleasure to aknowledge a very inspiring discussion with Prof. A. Abada.

\par 

\end{document}